\documentclass[english,conference,a4paper]{IEEEtran}
\usepackage[T1]{fontenc}
\usepackage{babel}
\usepackage{amsmath}
\usepackage{amsthm}
\usepackage{amssymb}
\usepackage{stmaryrd}
\usepackage{graphicx}
\usepackage{microtype}
\usepackage[unicode=true,
 bookmarks=false,
 breaklinks=false,pdfborder={0 0 1},backref=false,colorlinks=false]
 {hyperref}
\hypersetup{pdftitle={A Tight Converse to the Spectral Resolution Limits via Convex Programming},
 pdfauthor={Maxime Ferreira Da Costa and Wei Dai}}

\makeatletter
\theoremstyle{plain}
\newtheorem{thm}{\protect\theoremname}
\theoremstyle{plain}
\newtheorem{prop}[thm]{\protect\propositionname}
\theoremstyle{definition}
\newtheorem{defn}[thm]{\protect\definitionname}
\theoremstyle{plain}
\newtheorem{lem}[thm]{\protect\lemmaname}
\theoremstyle{plain}
\newtheorem{fact}[thm]{\protect\factname}

\usepackage[noadjust]{cite}

\makeatother

\providecommand{\definitionname}{Definition}
\providecommand{\factname}{Fact}
\providecommand{\lemmaname}{Lemma}
\providecommand{\propositionname}{Proposition}
\providecommand{\theoremname}{Theorem}

\begin{document}
\IEEEoverridecommandlockouts

\title{A Tight Converse to the Spectral Resolution Limit\\
via Convex Programming}

\author{\IEEEauthorblockN{Maxime Ferreira Da Costa and Wei Dai\thanks{This work was funded  by DSTL (Award DSTLX-1000118753), ONRG (Award N62909-16-1-2052) and by an EPSRC Doctoral Training Award.}}\IEEEauthorblockA{Department of Electrical and Electronic Engineering, Imperial College
London, United Kingdom\\
Email: \{maxime.ferreira, wei.dai1\}@imperial.ac.uk}}
\maketitle
\begin{abstract}
It is now well understood that convex programming can be used to estimate
the frequency components of a spectrally sparse signal from $2m+1$
uniform temporal measurements. It is conjectured that a phase transition
on the success of the total-variation regularization occurs when the
distance between the spectral components of the signal to estimate
crosses $1/m$. We prove the necessity part of this conjecture by
demonstrating that this regularization can fail whenever the spectral
distance of the signal of interest is asymptotically equal to $1/m$.
\end{abstract}

\global\long\def\trans{\mathsf{T}}

\global\long\def\herm{*}

\section{Introduction}

\subsection{Line spectral estimation}

Inferring on the fine scale properties of a signal from its coarse
measurements is a common signal processing problem that finds a myriad
of applications in various areas of applied and experimental sciences.
\emph{Line spectral estimation} is probably one of the most iconic
and fundamental instances of this category of problems, with direct
application in optics, radar systems, medical imaging and telecommunications.
In its most common formulation, it consists in recovering the locations
of highly localized patterns, or spikes, in the spectrum of a time
signal by observing a finite number of its uniform samples.

Denote by $\mathbb{T}=\left[0,1\right)$ the unidimensional torus
and let by $\mathcal{M\left(\mathbb{T}\right)}$ the set of complex-valued
Radon measures defined over $\mathbb{T}$. The line spectral estimation
problem aims to estimate the parameters of a sparse measure $\mu\in\mathcal{M}\left(\mathbb{T}\right)$
of the form 
\[
\forall\omega\in\mathbb{T},\quad\mu\left(\omega\right)=\sum_{k=1}^{s}c_{k}\delta_{x_{k}}\left(\omega\right)
\]
from its projection unto the first $2m+1$ complex trigonometric moments
$y\in\mathbb{C}^{2m+1}$ given by $y_{k}=\left\langle e^{i2\pi k\omega},\mu\right\rangle $
for $\left|k\right|\leq m$. In the above, the finite subset $X=\left\{ x_{k}\right\} _{k=1}^{s}\subset\mathbb{T}$
classically represents the support of the frequencies to estimate
and the subset $C=\left\{ c_{k}\right\} _{k=1}^{s}\subset\mathbb{C}$
represents its associated complex amplitudes. The sparse measure $\mu$
is assumed to be unknown, meaning that both $X$, $C$, and $s$ are
unknown parameters to estimate. As a result, the observation vector
$y=\left[y_{-m},\cdots,y_{m}\right]^{\trans}\in\mathbb{C}^{2m+1}$
has for integral representation
\begin{equation}
y=\int_{\mathbb{T}}a\left(\omega\right){\rm d\mu\left(\omega\right)},\label{eq:Consistency}
\end{equation}
whereby each ``atom'' $a\left(\cdot\right)\in\mathbb{C}^{2m+1}$
is the vector defined by $a\left(\omega\right)=\left[e^{-i2\pi m\omega},e^{-i2\pi\left(m-1\right)\omega},\cdots,e^{i2\pi m\omega}\right]^{\trans}$
for all $\omega\in\mathbb{T}$.

Recovering $\mu$ from the sole knowledge of $y$ is obviously an
ill-posed problem, since the set of measures $\mu\in\mathcal{M}\left(\mathbb{T}\right)$
leading to the same observation $y$ forms an affine subspace of $\mathcal{M}\left(\mathbb{T}\right)$
of uncountable dimension. The line spectral estimation problem aims
to recover the sparsest measure $\mu$ (the one of smallest support)
that is consistent with the measurement $y$ for the observation model
(\ref{eq:Consistency}). Hence the \emph{optimal estimator} can be
formulated as the output of the abstract optimization program
\begin{equation}
\mu_{0}=\arg\min_{\mu\in\mathcal{M}\left(\mathbb{T}\right)}\left\Vert \mu\right\Vert _{0},\;\text{subject to }y=\int_{\mathbb{T}}a\left(\omega\right){\rm d\mu\left(\omega\right)},\label{eq:xi0}
\end{equation}
whereby $\left\Vert \cdot\right\Vert _{0}$ is the pseudo-norm counting
the potentially infinite carnality of a complex Radon measure in $\mathcal{M}\left(\mathbb{T}\right)$.

\subsection{The spectral resolution limit}

By opposition to the ``classic'' finite-dimensional inverse problem
framework, one seeks, in the studied settings, to reconstruct \emph{continuously}
a subset $X$ over $\mathbb{T}$, instead of assuming that $X$ belongs
to a predefined finite subset of atoms. As a result, the notion of
restricted isometric property (RIP) or incoherence commonly used to
guarantee a robust inversion cannot be translated in the present problem
\cite{candes2006stable}. In particular, two atoms $a\left(\omega\right)$
and $a\left(\omega+\Delta\omega\right)$ will become more and more
coherent as $\Delta\omega$ tends to zero, and inferring on their
joined presence in the support set $X$ will become a harder and harder
task \cite{2040-8986-14-8-083001}. Hence, one can intuit that the
reconstruction performances of the support set $X$ are driven by
its minimal warp-around distance over the torus $\Delta_{\mathbb{T}}\left(X\right)$,
defined by
\[
\forall X\subseteq\mathbb{T},\quad\Delta_{\mathbb{T}}\left(X\right)\triangleq\inf_{\substack{x,x^{\prime}\in X\\
x\neq x^{\prime}
}
}\min_{p\in\mathbb{Z}}\left|x-x^{\prime}+p\right|.
\]

It was recently proven \cite{Moitra2014} that the line spectral estimation
problem is intractable whenever $\Delta_{\mathbb{T}}\left(X\right)<\frac{1}{m}$
in the sense that one can always find another discrete support set
$X^{\prime}\subset\mathbb{T}$ that can explain the observations $y$
within exponentially small noise levels with respect to the number
of measurements $m$. Hence, under this critical \emph{resolution
limit}, $X$ and $X^{\prime}$ are statistically indistinguishable
in the limit where $m\rightarrow\infty$, no matter the chosen estimator.
This result is explained by the presence of a phase transition on
the behaviors of the extremal singular values of Vandermonde matrices
with collapsing nodes around the unit circle. The interested reader
may refer to \cite{Aubel2017b} for a discussion and extensions of
this phenomenon. Moreover, it is particularly relevant to study those
results under the light of the early work of Slepian \cite{Slepian1978b},
who showed that no discrete time signal of length $2m+1$ can asymptotically
concentrate its energy in a spectral bandwidth narrower than $\frac{1}{m}$.

\subsection{Reconstruction via convex optimization}

There is a vast literature in signal processing on spectral deconvolution
methods. The MUSIC algorithm is probably the most popular one, with
well understood guarantees \cite{LIAO201633}.

In the recent years, a growing enthusiasm has been placed in tackling
the line spectral estimation problem though the lens of \emph{convex
optimization }after the pioneer work \cite{Candes2014a} demonstrated
that convex programming could recover any sparse measure having a
support verifying $\Delta_{\mathbb{T}}\left(X\right)\geq\frac{2}{m}$
in absence of noise and for sufficiently large values of $m$. The
authors' original idea consists in swapping the cardinality counting
pseudo-norm in (\ref{eq:xi0}) by the total mass $\left|\cdot\right|\left(\mathbb{T}\right)$
of the measure defined by $\left|\mu\right|\left(\mathbb{T}\right)=\int_{\mathbb{T}}{\rm d}\left|\mu\right|$
for every $\mu\in\mathcal{M}\left(\mathbb{T}\right)$, which can be
easily interpreted as an extension of the classic $\ell_{1}$-norm
to the set of Radon measures. The so-called \emph{total-variation}
(TV) regularization of the combinatorial Program (\ref{eq:xi0}) reads
\begin{equation}
\mu_{{\rm TV}}=\arg\min_{\mu\in\mathcal{M}\left(\mathbb{T}\right)}\left|\mu\right|\left(\mathbb{T}\right)\;\text{subject to }y=\int_{\mathbb{T}}a\left(\omega\right){\rm d\mu\left(\omega\right)},\label{eq:xiTV}
\end{equation}
which is a well-defined \emph{convex program} over $\mathcal{M}\left(\mathbb{T}\right)$.

The sufficient separation limit was later enhanced to $\frac{1.26}{m}$
in \cite{Fernandez-granda2015}. As suggested by simulation results
\cite{Tang2013}, the convex approach is conjectured to work asymptotically
in the regime $\Delta_{\mathbb{T}}\left(X\right)>\frac{1}{m}$. Performance
guarantees and stability of the reconstruction under white Gaussian
noise have been derived in \cite{Li2016,Bhaskar2013}. Line spectral
estimation is a canonical example of sparse inverse problems defined
over the set of measures, we refer the interested reader to \cite{Duval2015,DeCastro2012,Bendory2016511}
for more generic aspects and extensions of this theory.

\section{Main results}

\subsection{Spectral resolution limit of TV-regularization}

The generic TV-regularization framework is known to fail to reconstruct
\emph{complex-valued} (or \emph{signed}) Radon measures if certain
minimal separation criteria are not met. Necessary conditions were
given in \cite{7148951} for a wide range of inverse problems using
the compacity properties of the derivation operator over certain associated
dual spaces of functions.

Applying the generic result \cite{7148951} to the presented line
spectral estimation problem indicates that (\ref{eq:xiTV}) can fail
whenever $\Delta_{\mathbb{T}}\left(X\right)<\frac{1}{\pi m}$. The
best bound up-to-date was derived in \cite{Duval2015}, showing that
failure can append whenever $\Delta_{\mathbb{T}}\left(X\right)<\frac{1}{2m}$.
The proof relies on an argument on the decay rate of trigonometric
polynomials around their supremal values \cite{turan1946rational}.

This work focuses on tightening the \emph{necessary minimal separation}
$\Delta_{\mathbb{T}}\left(X\right)$ for the success of the TV-regularized
Program (\ref{eq:xiTV}). Theorem \ref{thm:MinimalCondition} proposes
an improvement of the previous results by showing the existence of
measures having a minimal separation asymptotically close to $\frac{1}{m}$
for which the convex approach fails. This tight result validates one
side of the conjecture on the achievable spectral resolution limit
through TV-regularization and constitutes a significant step toward
a complete understanding of the phase transition.

\begin{thm}
[Necessary separation for TV-regularization]\label{thm:MinimalCondition}For
every real $\delta>2$, there exists $M_{\delta}\in\mathbb{N}$, such
that for every $m\geq M_{\delta},$ there exists a set $X_{m}=\left\{ x_{k}^{\left(m\right)}\right\} _{k=1}^{s_{m}}\subset\mathbb{T}$
verifying $\Delta_{\mathbb{T}}\left(X_{m}\right)\geq\frac{1}{m}-\frac{\delta}{m^{2}}$
and a measure $\mu_{m}=\sum_{k=1}^{s_{m}}c_{k}^{\left(m\right)}\delta_{x_{k}^{\left(m\right)}}$
such that the solution of Program (\ref{eq:xiTV}) is not equal to
$\mu_{m}$.
\end{thm}
The demonstration of this result is provided in Section \ref{sec:Proof-of-the-main-result},
and is based on a construction of a specific sequence of measures
$\left\{ \mu_{m}\right\} _{m\in\mathbb{N}}$ for which we show the
\emph{non-existence} of associated dual certificates. To reach this
result, we introduce in Section \ref{sec:Proof-of-the-main-result}
the notion of \emph{stable diagonalizing families} of trigonometric
polynomials and highlight their relationships with the existence of
dual certificates. Theorem \ref{thm:Non-existence-of-diag-basis}
states that such families cannot exist if the support set is not separated
enough.

\subsection{Impact of the second order term}

\begin{figure}
\centering{}\includegraphics[bb=70bp 0bp 1003bp 564bp,clip,width=0.85\columnwidth]{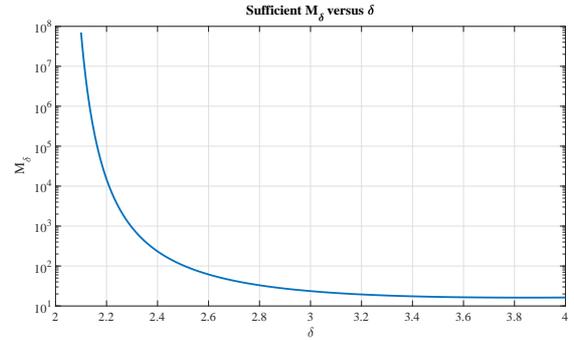}\caption{\label{fig:Mdelta}Upper bound on the minimal number of observations
$M_{\delta}$ requested by Theorem \ref{thm:MinimalCondition} against
the second order term $\delta$ . The curve admits a vertical asymptote
of equation $\log\left(M_{\delta}\right)=\Theta\left(\left(\delta-2\right)^{-1}\right)$
at $\delta\rightarrow2$.}
\end{figure}

Figure \ref{fig:Mdelta} presents \emph{sufficient} values of the
parameter $M_{\delta}$ defined in Theorem \ref{thm:MinimalCondition}
for different choices of the second order term $\delta$. Those results
are a by-product of the analysis (\ref{eq:L(m,d)}) in the proof of
Theorem \ref{thm:Non-existence-of-diag-basis}, and are presented
for illustration purposes. However, the present curve has a priori
no reason to act as a sharp bound on the minimal achievable value
of $M_{\delta}$.

\subsection{Notations}

Through this paper, $\left\llbracket s\right\rrbracket $ denotes
the sequence $\left[1,\cdots,s\right]$ for any $s\in\mathbb{N}$.
The set of $1$-periodic complex trigonometric polynomials of degree
$m$ is denoted $\mathcal{T}_{m}$, so that any element $Q\in\mathcal{T}_{m}$
writes for some vector $q\in\mathbb{C}^{2m+1}$ under the form
\[
\forall\omega\in\mathbb{T},\quad Q\left(\omega\right)=\sum_{k=-m}^{m}q_{k}e^{i2\pi k\omega}.
\]
The supremal norm over $\mathbb{T}$ is denoted $\left\Vert \cdot\right\Vert _{L_{\infty}}$.
For any $z\in\mathbb{C}^{*}$, the complex sign of $z$ is defined
by ${\rm sign}\left(z\right)=\frac{z}{\left|z\right|}$, and we let
by $\mathbb{U}=\left\{ z\in\mathbb{C}:\;\left|z\right|=1\right\} $
the complex unit circle.

\section{Proof of Theorem \ref{thm:MinimalCondition}\label{sec:Proof-of-the-main-result}}

\subsection{Dual certifiability}

It is now well understood that the success of TV-regularization methods
over the set of Radon measures is conditioned by the existence of
a so called \emph{dual certificate }\cite{Duval2015,Bendory2016511}:
A function representing the values of the optimal dual Lagrange variables
of Program (\ref{eq:xiTV}) and satisfying some extremal interpolation
properties. As a starting point of our analysis, we recall the following
proposition from \cite{Candes2014a}.
\begin{prop}
[Dual certificate]\label{prop:DualCertificate}The output of the
convex optimization program (\ref{eq:xiTV}) is equal to the ground
truth measure $\mu=\sum_{k=1}^{s}c_{k}\delta_{x_{k}}$ if and only
if there exists a complex trigonometric polynomial $Q\in\mathcal{T}_{m}$
satisfying
\begin{equation}
\begin{cases}
Q\left(x_{k}\right)={\rm sign}\left(c_{k}\right), & \forall k\in\left\llbracket s\right\rrbracket \\
\left|Q\left(\omega\right)\right|<1, & \forall\omega\notin X.
\end{cases}\label{eq:Dual-CertificateCondition}
\end{equation}
\end{prop}
We aim to demonstrate Theorem \ref{thm:MinimalCondition} by constructing
a sequence of well-separated measures $\left\{ \mu_{m}\right\} _{m\in\mathbb{N}}$
for which there is no element of $\mathcal{T}_{m}$ satisfying the
interpolation properties (\ref{eq:Dual-CertificateCondition}).

\subsection{Diagonalizing families}

In this subsection, we introduce the notion of diagonalizing families
over $\mathcal{T}_{m}$. Lemma \ref{lem:Diagonalizing-Lemma} draws
an important connection between the existence of dual certificate
for a measure $\mu$ and the existence of a stable diagonalizing family
on its support.
\begin{defn}
[Diagonalizing familly] Let $X=\left\{ x_{k}\right\} _{k=1}^{s}$
be a finite subset of $\mathbb{T}$. A \emph{first order diagonalizing
family} of $X$ over $\mathcal{T}_{m}$ is a set of $s$ elements
$\mathcal{P}_{X}=\left\{ P_{l}\right\} _{l=1}^{s}$ of $\mathcal{T}_{m}$
satisfying
\begin{equation}
\forall\left(l,k\right)\in\left\llbracket s\right\rrbracket ^{2},\quad\begin{cases}
P_{l}\left(x_{k}\right)=\delta_{l=k},\\
P_{l}^{\prime}\left(x_{k}\right)=0.
\end{cases}\label{eq:InterpProperties}
\end{equation}
\end{defn}
\begin{defn}
[Stable diagonalizing familly]\label{def:Stable-diag-familly} A
first order diagonalizing family $\mathcal{P}_{X}=\left\{ P_{l}\right\} _{l=1}^{s}$
of $X$ is said to be \emph{stable} if and only if $\left\Vert P_{l}\right\Vert _{L_{\infty}}=1$
for all $l\in\left\llbracket s\right\rrbracket $.
\end{defn}
\begin{lem}
\label{lem:Diagonalizing-Lemma}Let $X=\left\{ x_{k}\right\} _{k=1}^{s}$
be a discrete subset of $\mathbb{T}$ with cardinality $s\leq m$.
Suppose that for every $u\in\mathbb{U}^{s}$, there exists $Q_{u}\in\mathcal{T}_{m}$
such that
\begin{equation}
\begin{cases}
Q_{u}\left(x_{k}\right)=u_{k}, & \forall k\in\left\llbracket s\right\rrbracket \\
\left|Q_{u}\left(x\right)\right|<1, & \forall x\notin X,
\end{cases}\label{eq:InterpAssumption}
\end{equation}
then $X$ admits at least one first order stable diagonalizing family
over $\mathcal{T}_{m}$.
\end{lem}
\begin{IEEEproof}
Denote by $\mathcal{U}=\left\{ u^{\left(k\right)}\right\} _{k=1}^{s}$
the set of vectors defined by $u^{\left(k\right)}=\left[1,e^{i2\pi\left(k-1\right)},\cdots,e^{i2\pi\left(k-1\right)\left(s-1\right)}\right]^{\trans}\in\mathbb{U}^{s}$
for all $k\in\left\llbracket s\right\rrbracket $. By assumption there
exist $s$ polynomials $\mathcal{Q=}\left\{ Q_{u^{\left(k\right)}}\right\} _{k=1}^{s}$
satisfying Property (\ref{eq:InterpAssumption}) for each of the $u^{\left(k\right)}$.
Since $s\leq m$, the set of vectors $\mathcal{U}$ forms a basis
of $\mathbb{C}^{s}$. Hence, a classic interpolation theory argument
ensures that the set of trigonometric polynomials $\mathcal{Q}$ constitutes
a free family of $\mathcal{T}_{m}$, thus $\mathcal{Q}$ spans a sub-vectorial
space of $\mathcal{T}_{m}$ of dimension $s$.

We aim to build a stable diagonalizing family $\mathcal{P}_{X}$ of
$X$ lying in the span of the family $\mathcal{Q}$. Namely, we construct
\[
\forall l\in\left\llbracket s\right\rrbracket ,\quad P_{l}=\sum_{k=1}^{s}\alpha_{k}^{\left(l\right)}Q_{u^{\left(k\right)}},
\]
where $\left\{ \alpha^{\left(l\right)}\right\} _{l=1}^{s}\subset\mathbb{C}^{s}$
are coefficients to be determined. Each vector $\alpha^{\left(l\right)}$
is the unique solution of the linear system
\vspace{-1cm}
\begin{align*}
\forall l\in\left\llbracket s\right\rrbracket ,\quad\delta_{k=l}=P_{l}\left(x_{k}\right) & =\sum_{k=1}^{s}\alpha_{k}^{\left(l\right)}Q_{u^{\left(k\right)}}\left(x_{l}\right)\\
 & =\sum_{k=1}^{s}\alpha_{k}^{\left(l\right)}e^{i2\pi\left(k-1\right)\left(l-1\right)},
\end{align*}
that reformulates for every $l\in\left\llbracket s\right\rrbracket $
under the matrix form ${\bf F}_{s}\alpha^{\left(l\right)}=e_{l},$
whereby ${\bf F}_{s}\in\mathbb{M}_{s}\left(\mathbb{C}\right)$ is
the discrete Fourier transform matrix of dimension $s$, and $e_{l}$
denotes the $l^{\textrm{th}}$ vector of the canonical basis of $\mathbb{C}^{s}$.
${\bf F}_{s}$ is invertible with inverse ${\bf F}_{s}^{-1}=\frac{1}{s}{\bf F}_{s}^{\herm}$,
and consequently each polynomial $P_{l}$ reads
\begin{equation}
\forall l\in\left\llbracket s\right\rrbracket ,\quad P_{l}=\frac{1}{s}\sum_{k=1}^{s}e^{-i2\pi\left(k-1\right)\left(l-1\right)}Q_{u^{\left(k\right)}},\label{eq:Fl-linearCombinaison}
\end{equation}
and $\mathcal{P}_{X}$ verifies by construction the first condition
of (\ref{eq:InterpProperties}).

Next, since $\left|Q_{u}\left(x_{k}\right)\right|=\left|u_{k}\right|=1$
and $\left|Q_{u}\left(\omega\right)\right|<1$ for every element $\omega$
lying in a small open ball centered on $x_{k}$, one may conclude
that $Q_{u}^{\prime}\left(x_{k}\right)=0$ for all $k\in\left\llbracket s\right\rrbracket $.
Hence, by linearity, $P_{l}$ also satisfies $P_{l}^{\prime}\left(x_{k}\right)=0$
for all $k\in\left\llbracket s\right\rrbracket $. The second condition
of (\ref{eq:InterpProperties}) is verified and $\mathcal{P}_{X}$
is a first order diagonalizing family for $X$ over $\mathcal{T}_{m}$. 

Finally, one proves the stability of the family $\mathcal{P}_{X}$
by applying the triangular inequality to Equation (\ref{eq:Fl-linearCombinaison})
\[
\forall\omega\in\mathbb{T},\quad\left|P_{l}\left(\omega\right)\right|\leq\frac{1}{s}\sum_{k=1}^{s}\left|Q_{u^{\left(k\right)}}\left(\omega\right)\right|\leq1,
\]
which ensures that $\left\Vert P_{l}\right\Vert _{L_{\infty}}\leq1$.
Furthermore, since $\left|P_{l}\left(x_{l}\right)\right|=1$, one
has as well $\left\Vert P_{l}\right\Vert _{L_{\infty}}\geq1$ for
all $l\in\left\llbracket s\right\rrbracket $. Hence $\left\Vert P_{l}\right\Vert _{L_{\infty}}=1$,
and the stability property of $\mathcal{P}_{X}$ follows.
\end{IEEEproof}

\subsection{Existence of stable diagonalizing families}

It is worth noticing that, by a classic linear algebra argument, any
set $X\subset\mathbb{T}$ with cardinality $s\leq m$ admits infinitely
many diagonalizing families. However, the existence of a stable one
is not necessary guaranteed. Theorem \ref{thm:Non-existence-of-diag-basis}
states that there exist sets with asymptotic minimal distance $\frac{1}{m}$
that do not admit a stable diagonalizing family over $\mathcal{T}_{m}$.
Its demonstration is delayed to Section \ref{sec:Proof-of-Existence-Diag-Familly}
for readability.
\begin{thm}
\label{thm:Non-existence-of-diag-basis}For every real $\delta>2$,
there exists $M_{\delta}\in\mathbb{N}$, such that for every $m\geq M_{\delta},$
there exists a set $X_{m}=\left\{ x_{k}^{\left(m\right)}\right\} _{k=1}^{s_{m}}\subset\mathbb{T}$
such that $\Delta_{\mathbb{T}}\left(X_{m}\right)\geq\frac{1}{m}-\frac{\delta}{m^{2}}$
and there is no stable diagonalizing family of $X$ over $\mathcal{T}_{m}$.
\end{thm}

\subsection{Conclusion on Theorem \ref{thm:MinimalCondition}}

We now have all the elements to complete the proof of Theorem \ref{thm:MinimalCondition}.
Let $\delta>2$ and $m\in\mathbb{N}$ sufficiently large so that one
can pick a subset $X_{m}=\left\{ x_{k}\right\} _{k=1}^{s_{m}}\subset\mathbb{T}$
as in Theorem \ref{thm:Non-existence-of-diag-basis}. Using the contraposition
of Lemma \ref{lem:Diagonalizing-Lemma} on $X_{m}$, there must exist
one sign pattern $u\in\mathbb{U}^{s_{m}}$ such that there is no trigonometric
polynomial verifying Conditions (\ref{eq:Dual-CertificateCondition}).
Consider a measure $\mu_{m}$ of the form $\mu_{m}=\sum_{k=1}^{s_{m}}\tau_{k}u_{k}\delta_{x_{k}}$,
whereby $\left\{ \tau_{k}\right\} _{k=1}^{s_{m}}$ is a set of \emph{strictly
positive} reals. One has ${\rm sign}\left(\tau_{k}u_{k}\right)=u_{k}$,
and we conclude using the negation of Proposition \ref{prop:DualCertificate}
that the measure $\mu_{m}$ is not solution of Program (\ref{eq:xiTV}).\hfill{}\IEEEQED

\section{Proof of Theorem \ref{thm:Non-existence-of-diag-basis}\label{sec:Proof-of-Existence-Diag-Familly}}

Let $m\in\mathbb{N}$ , and let $X=\left\{ x_{k}\right\} _{k=1}^{s}$
be a subset of $\mathbb{T}$ with cardinality $s$. First of all,
if $P_{l}\in\mathcal{T}_{m}$ is the $l^{{\rm th}}$ element of a
diagonalizing family $\mathcal{P}_{X}$ over the set $X$, $P_{l}$
and its derivative $P_{l}^{\prime}$ both cancels by definition at
every point $x_{k}$ for $k\neq l$. Consequently $P_{l}$ has roots
with multiplicity two at each of those locations, and $P_{l}$ belongs
to the ideal generated by the minimal vanishing trigonometric polynomial
$Z_{X,l}\in\mathcal{T}_{s-1}$ defined by
\begin{equation}
\forall\omega\in\mathbb{T},\quad Z_{X,l}\left(\omega\right)\triangleq\prod_{\substack{1\leq k\leq s\\
k\neq l
}
}\frac{\sin^{2}\left(\pi\left(\omega-x_{k}\right)\right)}{\sin^{2}\left(\pi\left(x_{l}-x_{k}\right)\right)}.\label{eq:Zsquare}
\end{equation}
Hence, there exists a factorization of $P_{l}$ under the form 
\begin{equation}
\forall\omega\in\mathbb{T},\quad P_{l}\left(\omega\right)=Z_{X,l}\left(\omega\right)R_{l}\left(\omega\right),\label{eq:polynomZfactorization}
\end{equation}
where $R_{l}\in\mathcal{T}_{m-s+1}$. Using the assumptions on $P_{l}$
given by Equation (\ref{eq:InterpProperties}), the trigonometric
polynomial $R_{l}$ verifies the interpolation conditions
\begin{equation}
\begin{cases}
R_{l}\left(x_{l}\right)=P_{l}\left(x_{l}\right)=1\\
R_{l}^{\prime}\left(x_{l}\right)=\frac{P_{l}^{\prime}\left(x_{l}\right)-R_{l}\left(x_{l}\right)Z_{X,l}^{\prime}\left(x_{l}\right)}{Z_{X,l}\left(x_{l}\right)}=-\eta_{l},
\end{cases}\label{eq:Rl-InterpCondition}
\end{equation}
whereby we used the fact that $Z_{X,l}\left(x_{l}\right)=1$ and by
letting 
\[
\eta_{l}\triangleq Z_{X,l}^{\prime}\left(x_{l}\right)=2\pi\sum_{\substack{1\leq k\leq s\\
k\neq l
}
}\cot\left(\pi\left(x_{l}-x_{k}\right)\right).
\]

Next, we construct a well-separated subset of $\mathbb{T}$, and show
that no polynomial of the form (\ref{eq:polynomZfactorization}) is
stable in the sense of Definition \ref{def:Stable-diag-familly}.
For convenience, we restrict our analysis to odd trigonometric degrees
$m=2K+1$, and claim that the result is extendable for even values
of $m$. Let $\alpha_{m}\in\left(0,1\right)$ be such that $\frac{\alpha_{m}}{m+1}\triangleq\frac{1}{m}-\frac{\delta}{m^{2}}$
for some $\delta>1$ and consider a subset $X_{m,\delta}=\left\{ x_{k}^{\left(m,\delta\right)}\right\} _{k=-K}^{K}$
of $m$ equispaced elements of the form
\[
\forall k\in\left\llbracket -K,K\right\rrbracket ,\quad x_{k}^{\left(m,\delta\right)}\triangleq\frac{k\alpha_{m}}{m+1}.
\]
For every $m\in\mathbb{N}$, the minimal distance of $X_{m,\delta}$
reads
\[
\Delta_{\mathbb{T}}\left(X_{m,\delta}\right)=\frac{\alpha_{m}}{m+1}=\frac{1}{m}-\frac{\delta}{m^{2}}.
\]

Let $P_{0}\in\mathcal{T}_{m}$ be a diagonalizing polynomial of $X_{m,\delta}$
for the element $x_{0}^{\left(m\right)}=0$. $P_{0}$ can be factorized
(\ref{eq:polynomZfactorization}) under the form $P_{0}=Z_{m,\delta}\times R_{0}$,
where $Z_{m,\delta}\triangleq Z_{X_{m,\delta},0}\in\mathcal{T}_{m-1}$
is the minimal polynomial (\ref{eq:Zsquare}) that vanishes on $X_{m,\delta}\backslash\left\{ 0\right\} $
and $R_{0}\in\mathcal{T}_{1}$. By symmetry of $X_{m,\delta}$ around
$0$, $\eta_{0}=0$, and every trigonometric polynomial of degree
1 satisfying (\ref{eq:Rl-InterpCondition}) writes 
\begin{equation}
\forall\omega\in\mathbb{T},\quad R_{\gamma}\left(\omega\right)\triangleq\left(1-\gamma\right)+\gamma\cos\left(2\pi\omega\right),\label{eq:Rt-Def}
\end{equation}
for some $\gamma\in\mathbb{C}.$ Hence $P_{0}$ must have a factorization
of the form $P_{0}=P_{m,\delta,\gamma}\triangleq Z_{m,\delta}\times R_{\gamma}$
for some $\gamma\in\mathbb{C}$.

It remains to show that if $\alpha_{m}$ is small enough, every polynomial
of the form $P_{m,\delta,\gamma}$ verifies $\left\Vert P_{m,\delta,\gamma}\right\Vert _{L_{\infty}}>1$.
Formally, we aim to lower bound the quantity
\[
L\left(m,\delta\right)\triangleq\inf_{\gamma\in\mathbb{C}}\left\Vert P_{m,\delta,\gamma}\right\Vert _{L_{\infty}}=\inf_{\gamma\in\mathbb{C}}\sup_{\omega\in\mathbb{T}}\left|P_{m,\delta,\gamma}\left(\omega\right)\right|
\]
away from $1$ for small enough $\alpha_{m}$. Intuitively, we expect
$Z_{m,\delta}$ to reach large values far away from its roots, at
$\omega\simeq\frac{1}{2}$, and expect that the restrictive structure
(\ref{eq:Rt-Def}) on $R_{\gamma}$ will not leave the freedom to
drag the product $Z_{m,\delta}\left(\omega\right)R_{\gamma}\left(\omega\right)$
bellow $1$.

For ease of calculation, we introduce the translated polynomials $\tilde{Z}_{m,\delta}\left(\omega\right)=Z_{m,\delta}\left(\frac{1}{2}-\omega\right)$
and $\tilde{R}_{\gamma}\left(\omega\right)=R_{\gamma}\left(\frac{1}{2}-\omega\right)$
for all $\omega\in\mathbb{T}$, and let $\Omega_{m}=\left[-\frac{\alpha_{m}}{m+1},\frac{\alpha_{m}}{m+1}\right]\subset\mathbb{T}.$
The two following key lemmas, demonstrated in Section \ref{sec:ProofsCoreLemmas},
provide lower bounds on $\tilde{Z}_{m,\delta}\left(\omega\right)$
and $\tilde{R}_{\gamma}\left(\omega\right)$ over the set $\Omega_{m}$.
\begin{lem}
\label{lem:BoundZ0Square}There exists a constant $C\left(\delta\right)>0$
such that
\[
\forall m\in\mathbb{N},\forall\omega\in\Omega_{m},\;\tilde{Z}_{m,\delta}\left(\omega\right)\geq C\left(\delta\right)\left(m+1\right)^{2\left(\delta-1\right)}.
\]
\end{lem}
\begin{lem}
\label{lem:BoundR}Let $R_{\gamma}\in\mathcal{T}_{1}$ be has in (\ref{eq:Rt-Def}),
then 
\begin{equation}
\kappa_{m}\triangleq\inf_{\gamma\in\mathbb{C}}\sup_{\omega\in\Omega_{m}}\left|\tilde{R}_{\gamma}\left(\omega\right)\right|\geq\frac{\pi^{2}\alpha_{m}^{2}}{2\left(m+1\right)^{2}}.\label{eq:kappa-def}
\end{equation}
\end{lem}
One may lower bound the quantity $L\left(m,\delta\right)$ by controlling
the infimum of each of the factor of $P_{m,\delta,\gamma}$. Applying
Lemma \ref{lem:BoundZ0Square} and Lemma \ref{lem:BoundR} leads to
\begin{align}
L\left(m,\delta\right) & =\inf_{\gamma\in\mathbb{C}}\sup_{\omega\in\mathbb{T}}\left|Z_{m,\delta}\left(\omega\right)R_{\gamma}\left(\omega\right)\right|\nonumber \\
 & =\inf_{\gamma\in\mathbb{C}}\sup_{\omega\in\mathbb{T}}\left|\tilde{Z}_{m,\delta}\left(\omega\right)\tilde{R}_{\gamma}\left(\omega\right)\right|\nonumber \\
 & \geq\inf_{\gamma\in\mathbb{C}}\sup_{\omega\in\Omega_{m}}\left|\tilde{Z}_{m,\delta}\left(\omega\right)\tilde{R}_{\gamma}\left(\omega\right)\right|\nonumber \\
 & \geq\inf_{\omega\in\Omega_{m}}\tilde{Z}_{m,\delta}\left(\omega\right)\times\inf_{\gamma\in\mathbb{C}}\sup_{\omega\in\Omega_{m}}\left|\tilde{R}_{\gamma}\left(\omega\right)\right|\nonumber \\
 & =\frac{C\left(\delta\right)\pi^{2}\alpha_{m}^{2}}{2}\left(m+1\right)^{2\left(\delta-2\right)}=\Theta\left(m^{2\left(\delta-2\right)}\right).\label{eq:L(m,d)}
\end{align}
Hence, if $\delta>2$, $L\left(m,\delta\right)$ diverges when $m$
grows large. Consequently, there exists $M_{\delta}>0$ such that,
for all $m\geq M_{\delta}$, there is no stable diagonalizing family
of $X_{m,\delta}$ over $\mathcal{T}_{m}$.\hfill{}\IEEEQED

\section{Proofs of the auxiliary lemmas\label{sec:ProofsCoreLemmas}}

\subsection{Proof of Lemma \ref{lem:BoundZ0Square}: Lower bound on $Z_{0}\left(\omega\right)$}

The roots $\left\{ \tilde{x}_{k}^{\left(m,\delta\right)}\right\} _{\left|k\right|=1}^{K}$
of $\tilde{Z}_{m,\delta}$ are given by the relation $\tilde{x}_{k}^{\left(m,\delta\right)}=\frac{1}{2}-x_{K-k+1}^{\left(m,\delta\right)}$,
and a direct calculation yields
\begin{align*}
\forall k\in\left\llbracket K\right\rrbracket ,\quad\begin{cases}
\tilde{x}_{k}=\beta_{m}+\frac{k\alpha_{m}}{m+1}\\
\tilde{x}_{-k}=-\beta_{m}-\frac{k\alpha_{m}}{m+1}
\end{cases}
\end{align*}
whereby $\beta_{m}\triangleq\frac{1}{2}\left(1-\alpha_{m}\right)>0$
is an offset factor. Using Expression (\ref{eq:Zsquare}), one may
rearrange $\tilde{Z}_{m,\delta}$ as follows 
\begin{multline*}
\forall\omega\in\mathbb{T},\quad\tilde{Z}_{m,\delta}\left(\omega\right)=\\
\prod_{k=1}^{K}\frac{\sin^{2}\left(\pi\left(\beta_{m}+\frac{k\alpha_{m}}{m+1}-\omega\right)\right)\sin^{2}\left(\pi\left(\beta_{m}+\frac{k\alpha_{m}}{m+1}+\omega\right)\right)}{\sin^{4}\left(\pi\frac{k\alpha_{m}}{m+1}\right)}.
\end{multline*}
The polynomial $\tilde{Z}_{m,\delta}$ has no root over the set $\Omega_{m}$,
hence its logarithm $\tilde{z}_{m,\delta}$ is well defined over $\Omega_{m}$,
and it yields
\vspace{-0.15cm}
\begin{multline}
\hspace{-0.27cm}\forall\omega\in\Omega_{m},\,\tilde{z}_{m,\delta}\left(\omega\right)=\sum_{k=1}^{K}2\ln\sin\left(\pi\left(\beta_{m}+\frac{k\alpha_{m}}{m+1}-\omega\right)\right)\\
+2\ln\sin\left(\pi\left(\beta_{m}+\frac{k\alpha_{m}}{m+1}+\omega\right)\right)-4\ln\sin\left(\pi\frac{k\alpha_{m}}{m+1}\right).\label{eq:zlogExpression}
\end{multline}
We derive a lower bound on $\tilde{z}_{m,\delta}$ over $\Omega_{m}$
by using the two following results, whose elementary proofs have been
skipped. 
\begin{fact}
\label{claim:lnsinBound}For any $t,h\in\mathbb{R}^{+}$ such that
$t+h\leq\frac{\pi}{2}$, we have that
\[
\ln\sin\left(t+h\right)-\ln\sin\left(t\right)\geq h\cot\left(t\right)-\frac{h^{2}}{2}\csc^{2}\left(t\right).
\]
\end{fact}
\begin{fact}
\label{claim:sumRiemann}For all odd integer $m\in\mathbb{N}$ such
that $m=2K+1$, and all $\alpha\in\left(0,1\right)$, the following
inequalities hold,
\begin{align*}
\sum_{k=1}^{K}\cot\left(\frac{\pi k\alpha}{m+1}\right) & \geq\frac{m+1}{\pi\alpha}\ln\left(m+1\right)\\
\sum_{k=1}^{K}\csc^{2}\left(\frac{\pi k\alpha}{m+1}\right) & \leq\frac{2\left(m+1\right)^{2}}{\pi^{2}\alpha^{2}}.
\end{align*}
\end{fact}
Since $\pi\left(\frac{K\alpha_{m}}{m+1}+\beta_{m}+\left|\omega\right|\right)<\frac{\pi}{2}$
for all $\omega\in\Omega_{m}$, one can apply two times Fact \ref{claim:lnsinBound}
to each term of the sum (\ref{eq:zlogExpression}), yielding
\begin{align}
\tilde{z}_{m,\delta}\left(\omega\right)\geq & 4\pi\beta_{m}\sum_{k=1}^{K}\cot\left(\frac{\pi k\alpha_{m}}{m+1}\right)\nonumber \\
 & \qquad-2\pi^{2}\left(\beta_{m}^{2}+\omega^{2}\right)\sum_{k=1}^{K}\csc^{2}\left(\frac{\pi k\alpha_{m}}{m+1}\right)\nonumber \\
\geq & \frac{4\beta_{m}\left(m+1\right)}{\alpha_{m}}\ln\left(m+1\right)-4\left(1+\frac{\beta_{m}^{2}\left(m+1\right)^{2}}{\alpha_{m}^{2}}\right)\nonumber \\
\geq & 2\left(\delta-1\right)\ln\left(m+1\right)-4-4\left(\delta-1\right)^{2}\label{eq:logzBound}
\end{align}
where we made use of Fact \ref{claim:sumRiemann}, $\left|\omega\right|\leq\frac{\alpha_{m}}{m+1}$,
and noticing that $\frac{\delta-1}{2}\leq\frac{\beta_{m}\left(m+1\right)}{\alpha_{m}}\leq\delta-1$
for all $m\in\mathbb{N}$. Taking back the exponential in (\ref{eq:logzBound})
leads to the desired result for a constant $C\left(\delta\right)=e^{-4\left(1+\left(\delta-1\right)^{2}\right)}$.\hfill{}\IEEEQED

\subsection{Proof of Lemma \ref{lem:BoundR}: Lower bound on $R\left(\omega\right)$}

Let $\omega_{{\rm max}}\in\left[0,\frac{\pi}{2}\right]$ and $\Omega=\left[-\omega_{{\rm max}},\omega_{{\rm max}}\right]\subset\mathbb{T}$,
and define $c=\cos^{2}\left(\pi\omega_{{\rm max}}\right)\in\left[0,1\right]$
for convenience. We aim to find the value of $\gamma$ for which the
supremum of $\left|\tilde{R}_{\gamma}\left(\omega\right)\right|$
is minimal over $\Omega$. Noticing that $\left|\tilde{R}_{\gamma}\left(\omega\right)\right|^{2}=\left(1-2\left|\gamma\right|c\right)^{2}$,
the infimum in (\ref{eq:kappa-def}) is achieved for some positive
real $\gamma$, hence
\[
\kappa_{\Omega}\triangleq\inf_{\gamma\in\mathbb{C}}\sup_{\omega\in\Omega}\left|\tilde{R}_{t}\left(\omega\right)\right|=\inf_{\gamma\in\mathbb{\mathbb{R}}^{+}}\sup_{\omega\in\Omega}\left|\tilde{R}_{t}\left(\omega\right)\right|.
\]
Moreover, for a fixed value of $\gamma$, the symmetry of the function
$\left|\tilde{R}_{\gamma}\left(\omega\right)\right|$ and its monotonic
behaviors over $\left[0,\omega_{{\rm max}}\right]$ imply that the
supremum is reached either on $0$ or on $\omega_{{\rm max}}$, leading
to
\begin{align}
\sup_{\omega\in\Omega}\left|\tilde{R}_{t}\left(\omega\right)\right| & =\max\left\{ \left|\tilde{R}_{t}\left(\text{0}\right)\right|,\left|\tilde{R}_{t}\left(\omega_{{\rm max}}\right)\right|\right\} \nonumber \\
 & =\max\left\{ \left|1-2\gamma\right|,\left|1-2\gamma c\right|\right\} .\label{eq:supMax}
\end{align}
Define the auxiliary function $y$ over $\mathbb{R}^{+}$ as $y\left(\gamma\right)=\left(1-2\gamma\right)^{2}-\left(1-2\gamma c\right)^{2}$.
$y\left(\gamma\right)$ is positive whenever the maximum (\ref{eq:supMax})
is reached at $0$ and negative whenever it is reached at $\omega_{{\rm max}}$.
The auxiliary function is parabolic in $\gamma$ and we have
\[
y\left(\gamma\right)=\left(1-c^{2}\right)\gamma^{2}-\left(1-c\right)\gamma,
\]
which takes positive values for $\gamma\geq\frac{1}{1+c}$. Hence
\[
\sup_{\omega\in\Omega}\left|\tilde{R}_{\gamma}\left(\omega\right)\right|=\begin{cases}
\left|1-2\gamma\right| & \text{if }\gamma\geq\frac{1}{1+c}\\
\left|1-2\gamma c\right| & \text{otherwise},
\end{cases}
\]
is a piecewise monotonic function in $\gamma$. By similar argument,
\begin{align*}
\kappa_{\Omega} & =\min\left\{ \left|1-\frac{2}{1+c}\right|,\left|1-\frac{2c}{1+c}\right|\right\} \\
 & =\frac{1-c}{1+c}=\frac{\sin^{2}\left(\pi\omega_{{\rm max}}\right)}{1+\cos^{2}\left(\pi\omega_{{\rm max}}\right)}\geq\frac{\pi^{2}\omega_{{\rm max}}^{2}}{2}.
\end{align*}
One concludes on the lemma by letting $\omega_{{\rm max}}=\frac{\alpha_{m}}{m+1}$.\hfill{}\IEEEQED

\bibliographystyle{IEEEtran}
\bibliography{BibTeX}

\end{document}